\documentclass[letterpaper,12pt]{article}
\usepackage{chicago,graphicx}
\usepackage[margin=2cm,nohead]{geometry}
\usepackage[T1]{fontenc}
\usepackage[utopia]{mathdesign}
\usepackage{tikz}
\usetikzlibrary{arrows}
\def\reals{{\rm I}\!{\bf R}}

\title{Regression with Distance Matrices}
\author{Julian J.\ Faraway\footnote{Department of
Mathematical Sciences, University of Bath, BA2 7AY,
United Kingdom, \texttt{jjf23@bath.ac.uk}}}
\date{\today}

\begin{document}
\maketitle{}

\begin{abstract}
  Data types that lie in metric spaces but not in vector spaces are difficult to use within the usual regression setting, either as the response and/or a predictor. We represent the information in these variables using distance matrices which requires only the specification of a distance function. A low-dimensional representation of such distance matrices can be obtained using methods such as multidimensional scaling. Once these variables have been represented as scores, an internal model linking the predictors and the response can be developed using standard methods. We call scoring the transformation from a new observation to a score while backscoring is a method to represent a score as an observation in the data space. Both methods are essential for prediction and explanation. We illustrate the methodology for shape data, unregistered curve data and correlation matrices using motion capture data from an experiment to study the motion of children with cleft lip.
\end{abstract}


\section{Introduction}
\label{sec:introduction}

Regression methodology has been extended over the years to encompass a
wider range of variable types. Categorical predictors have been
represented using dummy variables and generalised linear modeling has
allowed additional response types. More recently, functional data have
been woven into the framework by representing functions using basis
function expansions with the resulting vectors of coefficients now
easily incorporated into the regression. But other types of data are
less amenable to regression modeling.

Data types such as unregistered curves, shapes, images, trees,
covariance matrices and so on are difficult to integrate into the
standard regression framework because such objects do not lie in a
natural vector space. In some cases, linearisations can be achieved
using basis function or tangent space representations but this is not
straightforward. It is easier to define distances between such objects
and hence define a metric space. Requiring only a metric space rather
than a vector space would allow us to use regression modeling for a
wider class of data types.  In this paper, we envisage situations
where the predictors and/or response are represented as distance
matrices. We will demonstrate how regression can be performed with
distance matrices. We show how predictions for new cases can be made
and how we may interpret the relationship between the predictors and
the response.

There is some prior work on the case where the predictors are expressed as a distance matrix. \shortciteN{cuadras1990distance} and subsequent papers present an approach to regression where the predictors are expressed as a distance matrix and principal coordinates analysis (PCO) is used to generate scores. The response is then regressed on these scores. The advantage of the approach is that categorical predictors can be incorporated into the distance calculation in a way that differs from the usual dummy variables approach. Sometimes a better fit may be obtained from this approach. The drawback is that some of the explanatory value of regression coefficients in the standard approach is lost.  Regression based on a distance matrix for the predictors can be seen in other approaches such as Gaussian Process Regression - at the heart of such methods there is a distance matrix of the predictors. See \shortciteN{rasmussengaussian} for more.

In other fields, the response is treated as a distance matrix. In Ecology, abundance matrices of species are sometimes converted to distance matrices because the standard Gaussian assumptions of MANOVA can not be justified.
See \shortciteN{mcardle2001fitting} who show how the standard partitioning of sums of squares in MANOVA is still possible with only knowledge of the distance matrix of the response.

In \shortciteN{legendre1994modeling} and \citeN{Lichstein2006}, a modeling approach where both the response and predictors are represented as distance matrices is presented. However, the method unrolls the matrices, column by column, into vectors and then proceeds with regression which does not respect the geometry of the problem. This would not allow the extrapolation and interpolation necessary for the interpretation.

Although several papers have appeared treating either the response or the predictors as a distance matrix, the focus has been on hypothesis testing in the response case and prediction in the predictor case. Our focus here is the explanatory value of regression in showing which aspects of the response depend on which aspects of the predictors. Certainly there are many ``black box'' methods that could be applied to some of the situations we shall describe but even if these achieve some measure of predictive performance, this is of little value if explanation is the main goal of the model.

Geodesic regression is a different approach to modeling manifold-valued variables that has been developed in several papers such as
\shortciteN{fletcher2011geodesic} and \shortciteN{niethammer2011geodesic}. The approach we take here is more generic in nature. We aim to provide a method that can be used across a wide class of difficult data types. Obtaining general theoretical results that hold across such a wide class of objects would be challenging. Instead, this paper takes a case study approach to showing how interesting and practically useful results can be obtained while acknowledging that much rigorous work remains to be done.

In Section~\ref{sec:methods}, we describe the method in general. In Section~\ref{sec:shapes-shapes}, we discuss an example with shape variables as both predictor and response. In Section~\ref{sec:shapes-curves}, we demonstrate the modeling of an unregistered curve predictor and a shape response. Section~\ref{sec:correlation-matrices} covers an example involving a correlation matrix as a reponse. The data in all three cases come from a study of children with cleft lip where the objective is to understand factors that affect the motion of the face. In Section~\ref{sec:conclusion}, we end with some conclusions.

\section{Methods}
\label{sec:methods}

The proposed modeling process is illustrated in Figure~\ref{fig:modproc}. The predictors $X$ will be expressed through a distance matrix $D^X$. Low dimensional coordinates $S^X$ will be used to represent $D^X$. A similar process will be used to form $D^Y$ and $S^Y$ from the response $Y$. An internal model will be used to link $S^X$ and $S^Y$. By itself, this internal model can only measure the strength of the relationship between $X$ and $Y$. To make the method useful in practice, we describe how to map between the spaces of $X$ and $S^X$ and similarly for $Y$ and $S^Y$. This enables interpretation and prediction on the scales of $X$ and $Y$.

\tikzstyle{int}=[draw, fill=gray!20, minimum size=2em]

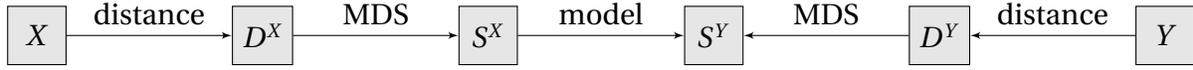
\begin{figure}[hbtp]
  \centering

\begin{tikzpicture}[node distance=3cm,auto,>=latex']
    \node (a) [int]{$X$};
    \node (b) [int][right of=a]{$D^X$};
    \node (c) [int][right of=b]{$S^X$};
    \node (d) [int][right of=c]{$S^Y$};
    \node (e) [int][right of=d]{$D^Y$};
    \node (f) [int][right of=e]{$Y$};
    \path[->] (a) edge node {distance}  (b);
    \path[->] (b) edge node {MDS} (c);
    \path[->] (c) edge node {model} (d);
    \path[<-] (d) edge node {MDS}(e);
    \path[<-] (e)  edge node{distance} (f);
\end{tikzpicture}

  \caption{Predictors $X$ form distance matrix $D^X$ from which scores $S^X$ are calculated using MDS. Similarly for the response $Y$. The scores are related using an internal model.
  \label{fig:modproc}}
\end{figure}

\subsection{Distance matrices and scores}
\label{sec:dist-matr-scor}

Consider a metric space $(M_X, d_x)$ where distance $d_x$ is a function such that $d_x : M_x \times M_x \rightarrow \reals$.  The choice of distance is critical to the outcome and the user must choose carefully according to the nature of $X$. Finding a suitable distance may not be easy, particularly on difficult manifolds.  Given a sample of data $x_1, \dots ,x_n$, we compute an $n \times n$ distance matrix $D^x$ where $D_{ij}^x = d_x(x_i,x_j)$. Similarly, we have another, and possibly quite different, metric space for the response, $(M_Y,d_y)$ and associated distance matrix $D^y$.

Classical multidimensional scaling (cMDS) forms a matrix $B$ from distance
matrix $D$ such that:
\begin{displaymath}
  B_{ij} = -(D^2_{ij} -D^2_{i\cdot}-D^2_{\cdot j} + D^2_{\cdot\cdot})/2
\end{displaymath}
where the dots in the subscripts indicate that means are taken over the index.  We form the eigendecomposition: $B=SS^T$ with eigenvalues $\lambda_1 \ge \lambda_2 \ge \dots \lambda_n$. The columns of matrix $S$ contain the principal coordinates or scores. We perform this decomposition on both $D^x$ and $D^y$ to obtain $S^x$ and $S^y$ respectively. There is some choice regarding the dimension of $S^x$ and $S^y$. Since there will be some later dimension reduction in the modeling, we should be inclusive in this selection. Nevertheless, the computational burden can be reduced by eliminating dimensions of the scores which correspond to relatively small eigenvalues.

There are many other ways of forming low-dimensional coordinate
representations of distance matrices, for example ISOMAP
(\shortciteN{tenenbaum00}), that might, in principle, be used
here. However, the calculations using just cMDS are quite complex so
the use of a more sophisticated method is not practical without some
advance on the methodology that we shall shortly describe.

\subsection{Internal Model}
\label{sec:model}

We need to model the relationship between $S^x$ and $S^y$. Since both
of these will usually be matrices, we need a regression-like method
that can handle a multivariate response. We also need some model
selection methods because we may obtain better results by dimension
reduction. We also require smoothness in that small changes in the
input should not result in large changes in the output.

Our particular choice of method is Partial Least Squares (PLS) using
an implementation based on \shortciteN{mevik2007pls} using the SIMPLS
method of \shortciteN{dejong:93}. PLS finds the linear combinations of
$S_X$ that are most strongly related to $S_Y$. PLS has
well-established variable selection methods based on cross-validation
which determine the size of the model that will give the best
predictive behaviour.

Various alternatives might be considered for modeling the relationship between the scores. Multivariate Multiple Regression, Principal Components Regression and Canonical Correlation analysis are all plausible alternatives while more modern methods from Machine Learning are also potential choices. There is no requirement that the internal model be interpretable because we plan to explain the relationship in terms of the original predictor and response spaces. Furthermore, a non-linear model could be used provided the fit is stable and the relationship smooth.

We like to use PLS here because it has well-developed methods of model
selection and produces a stable linear relationship.  Nevertheless,
there are many other methods that could be substituted here and we
claim no general superiority for PLS nor do we believe there is a
uniformly best choice of method here. If there is ambivalence, two
different methods could be used with the hope of confirming that the
conclusions are not sensitive to this choice.

\subsection{Scoring and Backscoring}
\label{sec:scoring-backscoring}

Suppose a predicted response is required for a new value of $x$. Following the modeling process shown in Figure~\ref{fig:modproc}, we need to convert this $x$ into a score (which we call \emph{scoring}). The internal  model will then be used to predict a response score. This score will then need to be converted back to the space of the response (which we call \emph{backscoring}). Scoring and backscoring are essential for the interpretation of the model in the spaces of the observed variables. This will usually require scoring and backscoring for both the predictors and the response.

Consider a new data point $x_{new}$ which we must map into the score space. Appending $x_{new}$ to the observed $x$'s and recomputing the distance matrix will change all the previously computed scores which is not sensible. We should maintain the existing scores while locating the new score that reflects the distances of $x_{new}$ to the observed $x$'s.
\shortciteN{gower1968apv} describes one way this may be achieved:
\begin{equation}
  \label{eq:scoring}
  s_{new}^x = \Lambda^{-1}S^T(d^2(\mathbf{x},\bar x) - d^2(\mathbf{x}, x_{new}))/2=\phi(x_{new})
\end{equation}
where $\Lambda =\textrm{diag}(\lambda_1, \dots , \lambda_n)$ and $\bar x$ is a centroid. However, $d^2(\mathbf{x},\bar x) = -diag(B)$ so explicit computation of the centroid is not yet needed. This computation requires only the distances $d(\mathbf{x},x_{new})$ of the new point to the original points. In practice, we retain only the first few coordinate directions so appropriately reduced versions of $\Lambda$ and $S$ would be used, and $s_{new}$ represents a projection onto the reduced score space. Other methods of scoring based on various motivations have been proposed. As discussed in \citeN{faraway2012backscoring}, several of these methods are equivalent to Gower's method.

Backscoring is more problematic. The internal linear model will produce $s_{new}^y$ which we want to map to the space of the observed response.  Each $s_{new}^y$ will correspond to a set of values in $M_y$ which have this score. Within this set, a natural choice is the one closest to the centroid, that is:
\begin{equation}
  \label{eq:backscoring}
  y_{new} = \mathrm{arg}\min_{s_{new}^y=\phi(y)} d(y,\bar y)
\end{equation}
This does require the computation of a centroid in the response space for which we may use  the Fr\'echet median:
\begin{displaymath}
  \mathrm{argmin}_{y \in M_y} \sum_{i=1}^n d(y,y_i)
\end{displaymath}

The choice of $y_{new}$ as the member of the set closest to the centroid can also be justified by analogy to Principal Components Analysis (PCA). Suppose we ask which point in the data space corresponds to a score where we specify only the first few components of that score because we have made some dimension reduction by discarding the remaining components. There is no unique answer because different values for the discarded scores would result in different values in the data space. But it is common practice to set the discarded scores to zero and thus obtain a unique solution. This solution also happens to be the member of the set closest to the mean. Hence our proposed solution is consistent with common practice in PCA-based solutions.

There are also questions regarding whether the feasible set is closed or compact and whether a unique minimiser exists. We do not believe a general answer is possible although some progress might be made by restricting attention to specific data types. Such an investigation would be extensive and lies beyond the scope of this paper.

Finding a solution to (\ref{eq:backscoring}) is difficult when the data do not lie in a vector space because most optimisation methods need this property. In a Euclidean space, the MDS is effectively equivalent to a PCA and the solution is explicit. But there would be little incentive to use our method for Euclidean spaces as more direct methods apply. The method proposed here is only valuable for the more difficult types of data for which the optimisation is problematic. The constraint that $y \in M_y$ further complicates the search for a solution which needs to be customised to the $M_y$. We will describe such methods in the three examples to follow.  Further discussion and examples of scoring and backscoring may be found in \shortciteN{faraway2012backscoring}.

\subsection{Adding Predictors}
\label{stec:adding-predictors}

Suppose we wish to add a predictor $Z$ to the set of predictors lying in $M_x$. To maintain the spirit of the approach explained above, we might seek to modify the distance measure $d_x$ to $d_{x,z}$ to accomodate the additional predictors.
However, when $X$ and $Z$ are quite different in nature, the construction of an appropriate distance would be difficult.
 Alternatively, we may have a distance measure on $Z$ and be able to obtain a distance matrix which might be combined with $D_x$ using
\begin{displaymath}
  D_{combined} = \sqrt ( D_x^2 + D_Z^2)
\end{displaymath}
However, there is the problem of the relative weighting of $X$ and $Z$ in the combined distance measure which in ordinary regression would be represented by a parameter. This is difficult to incorporate here so a different approach is recommended.

Consider the case where $Z$ comprises the usual quantitative and/or categorical predictors. We can use the standard regression methods to produce a model matrix $S^Z$ which can be adjoined to $S^X$ and the internal modeling can then proceed. A more difficult case arises where $Z$, like $X$ also lies in a non-vector space and we must resort to a distance matrix $D^z$. An MDS then produces a matrix of scores $S^Z$ which can then be adjoined to $S^X$. Difficulties may arise in the interpretation of the fitted model since we must distinguish the effect of $X$ from the effect of $Z$. But these problems arise also in standard regression situations so there is no additional difficulty in this setting.

\subsection{Inference and Diagnostics}
\label{sec:inference}

Inference and model checking can be made using the internal linear model. For most such methods, there are well-established methods of making this inference and we have nothing to add to this. However, we may wish to assess and check the fit in the scale of the response. We have residuals given by $d_y(\hat y_i , y_i)$ which may be summed to form a residual sum of squares (RSS).  Suppose we wish to compare a larger model with a smaller model, containing a subset of the predictors, then we can compute an F-statistic of the form:
\begin{displaymath}
  F = (RSS_{small} - RSS_{large})/RSS_{large}
\end{displaymath}

The significance of this F-statistic can be explored numerically using permutation tests which we demonstrate in the examples to come. A full permutation test requires that we compute the test statistc for all possible permutations of a predictor or the response, depending on what is being tested. The $p$-value is given by the fraction of permuted test statistics that exceed the observed value. For larger datasets, all permutations may take too long to compute. In this case, a large random sample of permutations can be used to adequately estimate the $p$-value.

We can also compute an $R^2$-like statistic of the form $1-RSS/TSS$ where $TSS$ is the total sum of squares derived from a null model of a constant predictor. The residuals may also be used to make diagnostics analogous to those used in common regression.

\section{Shapes on Shapes}
\label{sec:shapes-shapes}

\begin{figure}[hbtp]
  \centering
  \includegraphics{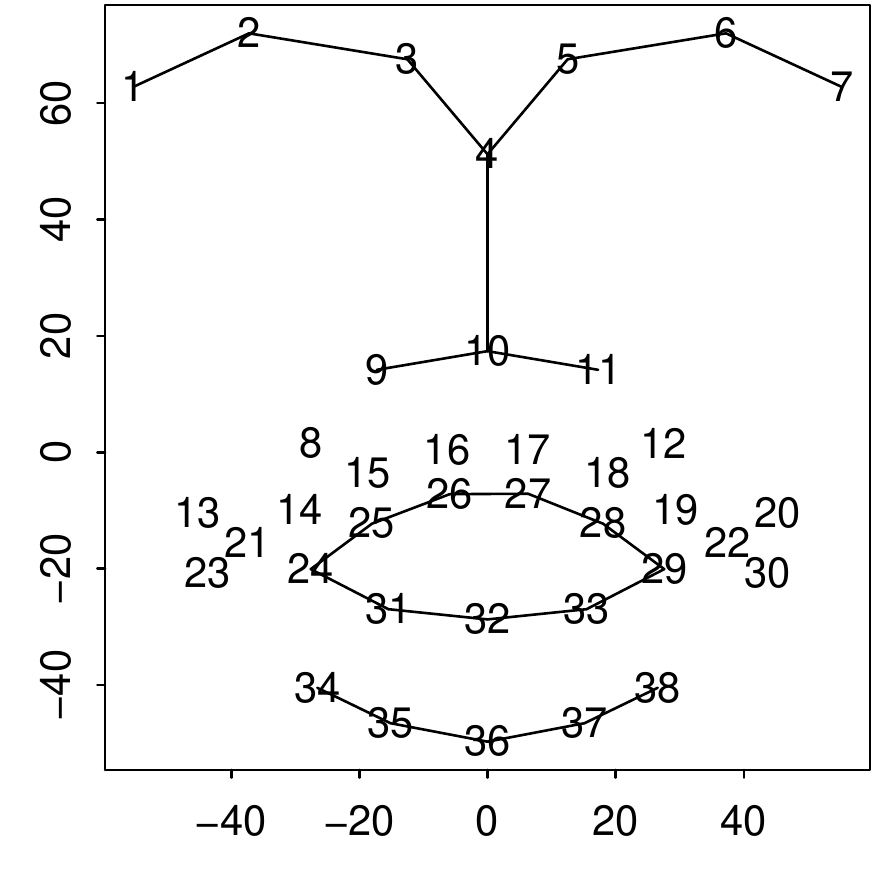}
  \caption{Configuration of 38 markers on the average face. Lines depicting the eyebrows, nose, mouth and lower jaw have been drawn.
  \label{fig:stdface}}
\end{figure}

We consider shape as the information in a configuration of landmarks that is invariant to rotation, translation and size transformations. The data for the example comes from a subset of a study described in \shortciteN{trotman10:_effec_of_lip_revis_surger}. 38 markers are placed in designated positions on the faces of 36 normal children as shown in Figure~\ref{fig:stdface} and motions in 3D are recorded for four second segments at 60Hz producing 240 frames of motion. Among other actions, the children are asked to smile (starting from and returning to rest). We will generate three different datasets from this experiment.

For the example in this section, we consider only the initial pose, defined by the average position during the first 10 frames of motion and the maximal pose, defined by the frame of motion which is furthest from the initial pose in terms of Procrustes distance. We consider the problem of relating the maximal pose described by a $38 \times 3$ shape matrix which lies in $\Sigma^{38}_3$ to the initial pose also described by a $38 \times 3$ shape matrix also lying in $\Sigma^{38}_3$. We are interested in whether the rest pose of the face has any effect on the type of smile subsequently made.

We construct the distance matrices using the Procrustes distance and then use cMDS to construct the scores as described in  Section~\ref{sec:dist-matr-scor}. These scores are approximately equivalent to the tangent space representation, provided the shapes are relatively concentrated. The distance and tangent space approximation are described in \citeN{dryden:98} with numerical implementation being provided by an R package called \texttt{shapes} from \shortciteN{dryden09:_shapes}.  Faces do not change shape that much during motion so the shapes are concentrated in this instance. This makes the necessary mappings between the score and data spaces very much easier to compute because the approximate linear mappings can reasonably be used. If the shapes were more dispersed or we chose another distance function, as explored in \citeN{faraway2012backscoring}, the backscoring becomes more difficult.

We consider the percentage of the variation explained by choosing the dimension of the score space. For both the predictor and the response, this variation drops off relatively slowly with only 20-25\% of the variation explained by the first component. We choose 10 dimensions in both spaces which ensures that no unchosen component explains more than 2-3\% of the variation.

In this application, our interest lies in explanation rather than prediction. For brevity in the exposition, we will focus on the first component of the response fitted by the PLS model. At least the second component of the response is of interest and could be pursued. Once we focus on the first dimension of the response, we find that, using leave-out-one crossvalidation, only one linear combination of the predictor scores is sufficient explanation. There is a correlation of 0.78 between these predictor and response scores indicating a strongly significant relationship. However, given the selection effect of choosing the maximally correlated linear combinations, we must be cautious about assuming this implies a strongly significant relationship between the shape predictors and response.

\begin{figure}[hbtp]
  \centering
  \includegraphics{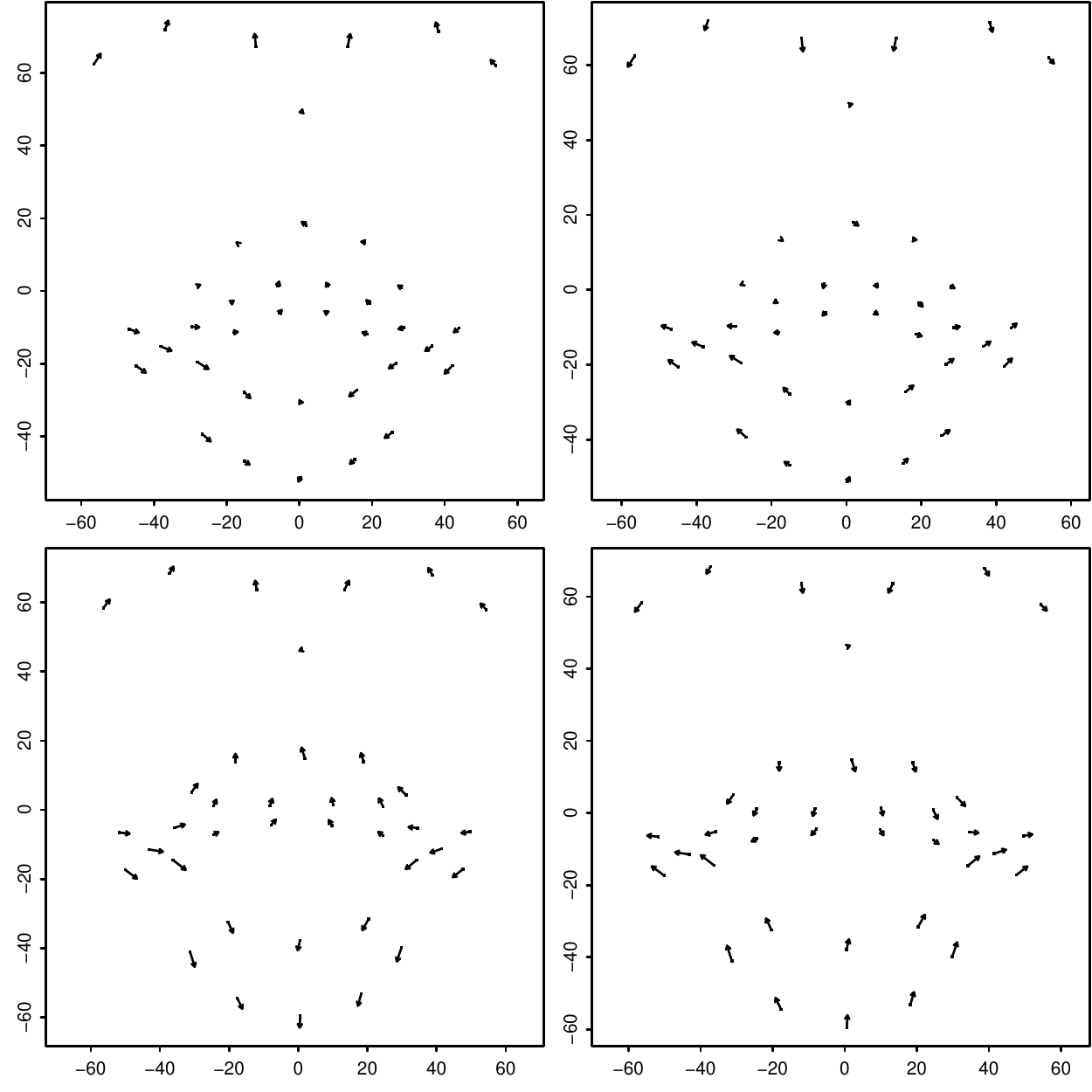}
  \caption{Arrows for each facial marker start from the mean pose and end perturbed by two SDs. The upper two plots show perturbations on the initial pose while the two lower plots show perturbations on the maximal pose. The plots on the left show perturbations up from the mean --- we see a pursed lip initial pose leads to a mouth-open smile. The plots on the right show perturbations down from the mean --- we see a thin grin initial pose leads to a mouth-closed smile.
  \label{fig:combface}}
\end{figure}

The scores can be mapped back to shapes using the appropriate linear combinations in the tangent space. These are depicted in Figure~\ref{fig:combface}.  In the top panel on the left, each arrow starts from the mean initial pose (corresponding to the markers in Figure~\ref{fig:stdface}) and ends at the location of that marker when perturbed up by two standard deviations in the direction of the first component. We can see this perturbation from the mean face corresponds to a pose where the lips are drawn in towards a kiss-like pose. On the top right, the contrasting perturbation down where we see the lips are drawn out into a thin grin. Note that motion is not being depicted in these plots, rather the variation about the mean pose and that the distinction between perturbing up and down is arbitrary.

These input scores can be related to predicted output scores which can again be represented in the shape space of the response as seen in the lower two plots of Figure~\ref{fig:combface}. Note that the principal coordinate decomposition only reliably defines axes of variation so ``up'' and ``down'' are arbitrary. Even so, once the two PCOs have been made, we can take care to be consistent in the interpretation of the direction hence the left panel above does correspond to the left panel below. In the lower left panel, we see that this direction of variation corresponds to a mouth open smile where the teeth would likely be visible while on the lower right, we see a closed mouth smile where the lips are kept together while the corners of the mouth move outward and upward. Thus the analysis shows that an initial pose closer to a kiss leads to a mouth open smile while a thin grin leads to a broader grin. Hence we see that the initial pose indicates some anticipation of the type of smile to come.

We can compute a test of significance for the predictor in this regression. The test statistic has a value of 0.179. Using a permutation test (where the response scores are permuted), we determine that the p-value is very small (none of the 1000 permuted test-statistics exceeded the observed value).

In this example, the scoring and backscoring were relatively easy to compute because of the good approximation of MDS to the tangent space approach for concentrated shapes. So unless we have shapes that are more dispersed or we choose another distance function, we do not need our approach. Nevertheless, it is worth showing that the method produces results consistent with a more established approach and that the possibility exists to tackle problems for which the tangent space approximation would not work. In the next example, the backscoring becomes more difficult.

\section{Shapes on Curves}
\label{sec:shapes-curves}

There is a growing literature on functional data analysis. Substantial work has been done to integrate one-dimensional functions both as predictors and responses in a regression problem. The usual approach is to approximate the functions as linear combinations of basis functions. The coefficients of these linear combinations are vectors which can be readily integrated into the standard regression framework. See \citeN{ramsay:97} for an overview. A common preliminary step in a functional data analysis is registration where phase variation in the observed curves is removed. In some cases, removing this variation is appropriate because it contains no information. We can also also model the phase variation seperately.
In the analysis to follow, we retain both the phase and amplitude variation in the observed curves as both forms of variation may jointly be of interest. There are other ways to do this joint modeling of phase and amplitude, such as \shortciteN{srivastava2011registration}, but these differ from the approach taken here.

\begin{figure}[hbtp]
  \centering
  \includegraphics{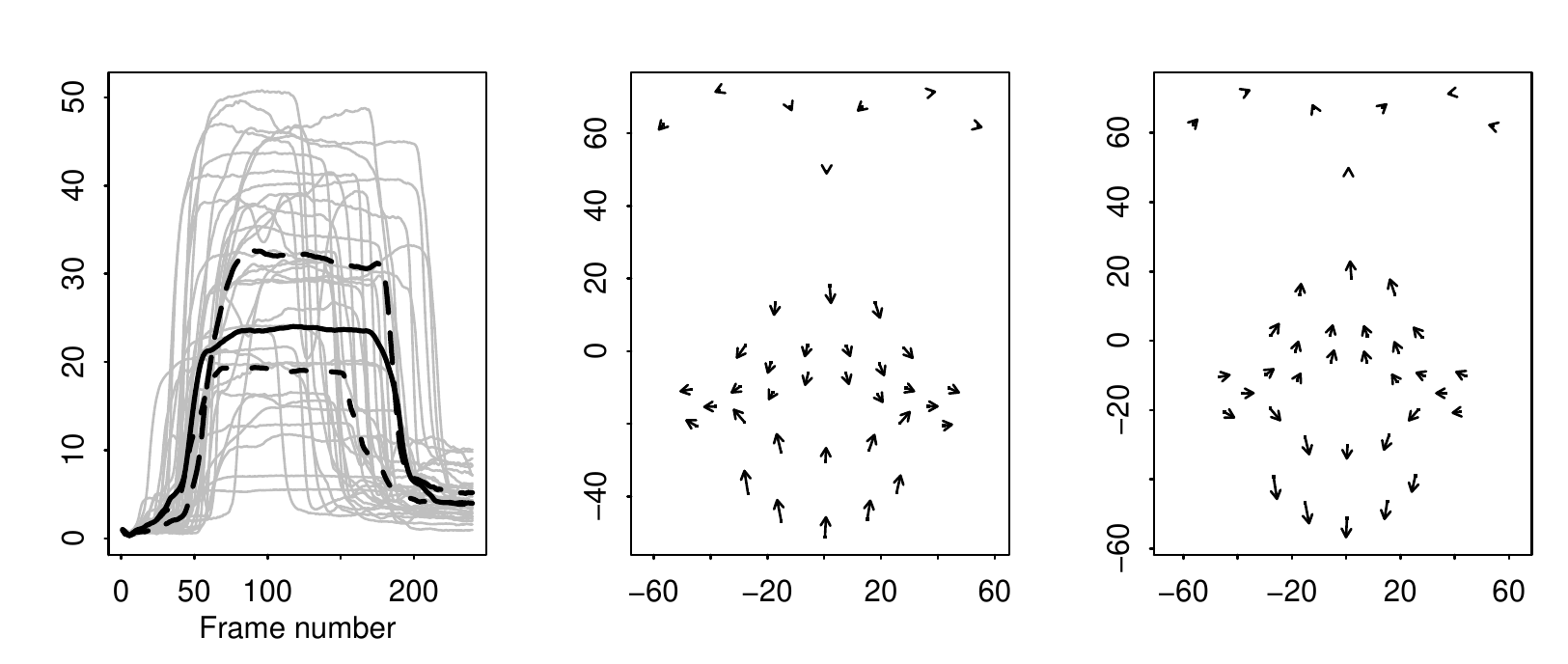}
  \caption{First plot shows the first component of the curve predictor variation. The data curves are shown in gray, the centroid as a thick black curve and the dashed lines show one sd variations around this mean in the axis of the first component. The second plot shows the variation from the mean shape corresponding to the lower of the two dashed curves in the first plot. The third plot shows the variation around the mean shape corresponding to the higher of the two dashed curves in the first plot.
  \label{fig:funcshap1}}
\end{figure}

Consider the same experiment as in the previous example except that we generate the data differently. We compute the Procrustes distance of each frame in the motion to the initial shape. This produces a set of curves which will be used as the predictor in this example and are shown as the gray curves in the first panel of Figure~\ref{fig:funcshap1}. The curve represents both the timing (phase variation) and the magnitude (amplitude variation) for the motion. We use the maximal pose, achieved at the greatest distance from rest as the response. We standardise this response as
\begin{displaymath}
  s_{std} = s_{max} - s_{initial} + s_{mean}
\end{displaymath}
where the shapes $s$ (respectively, standardized, maximal pose, initial pose, mean (over subjects) pose) are represented in a tangent space where addition and subtraction are meaningful. The purpose of this transformation is to remove the variation in the rest poses of the faces by estimating the maximal pose that would be achieved if started from the mean initial face. Hence the predictor is a curve and the response is a shape.

We use the Fr\'echet distance between curves defined in two dimensions: $(f(t),t)$. For comparability, we scale in both dimensions so that $f(t), t \in [0,1]$ by dividing by the maximum value of $f(t)$ across all the curves in the data. Now consider moving along one curve at rate $\alpha(s)$ and another curve at rate $\beta(s)$. The Fr\'echet distance is defined as
\begin{displaymath}
  d(f_1,f_2) = \inf_{\alpha,\beta} \max_{s \in [0,1]} \| (f_1(\alpha(s),\alpha(s)))-(f_2(\beta(s)),\beta(s)) \|
\end{displaymath}
where $\|\cdot\|$ denotes Euclidean distance between the pairs. Fr\'echet distance is sometimes called dog-leash distance since if a man walks along the first curve and his dog along the second, the distance is the shortest leash that could be used to enable them to traverse each curve (without backtracking). Because both functions are represented by discrete approximations, the distance can be computed efficiently.

The usual vector space of integrable functions is not sensible for use here. Consider the average $(f_1+f_2)/2$. If $f_1$ represents a short duration smile occuring near the beginning of the time period while $f_2$ is a short duration smile occuring near the end, the average will consist of two smaller smiles (which the subjects were not instructed to do). Intuitively, the best average would be a short duration smile in the middle of the time period but defining a vector space to achieve such an outcome is problematic.

It is very difficult to explicitly define the manifold of
curves that might represent the timing and magnitude of a smile. It is
apparent from examining the data that such curves have a common shape
but they are sufficiently irregular to resist precise
definition. Instead, we take a constructive approach to defining the
manifold. We observe that we know 36 members of the manifold. We
define a parameterized transformation of a curve and a way to compute
a weighted combination of two curves. The outputs are also assumed to
lie in the manifold as the transformation and combination are defined
in such a way as to preserve the qualitative properties of a smile
curve.  Of course, it is undesirable to define such a manifold in
terms of the observed data but a more extrinsic definition is
difficult to achieve.

We define a family of transformations of a curve $f$.
Consider mappings of the form: $f(t) \rightarrow f^p(t,\mathbf{p})$:
\begin{displaymath}
  f^p(t,\mathbf{p}) = \left\{
    \begin{array}{ll}
      p_3f(tp_1/p_2) & t \leq p_2 \\
      p_3f((1-p_1)t/(1-p_2) + (p_1-p_2)/(1-p_2)) & t > p_2
    \end{array} \right.
\end{displaymath}
where $p_1,p_2 \in [0,1]$ and $p_3 > 0$.  In words, pick a point $p_1$ in the interior and move it to $p_2$, linearly rescaling either side. Scale by $p_3$. An example of a transformation is shown in the first panel of Figure~\ref{fig:combface}, where the solid curve is transformed to the dashed curve using $p_1=0.4, p_2=0.6, p_3=0.8$.

\begin{figure}[hbtp]
  \centering
  \includegraphics{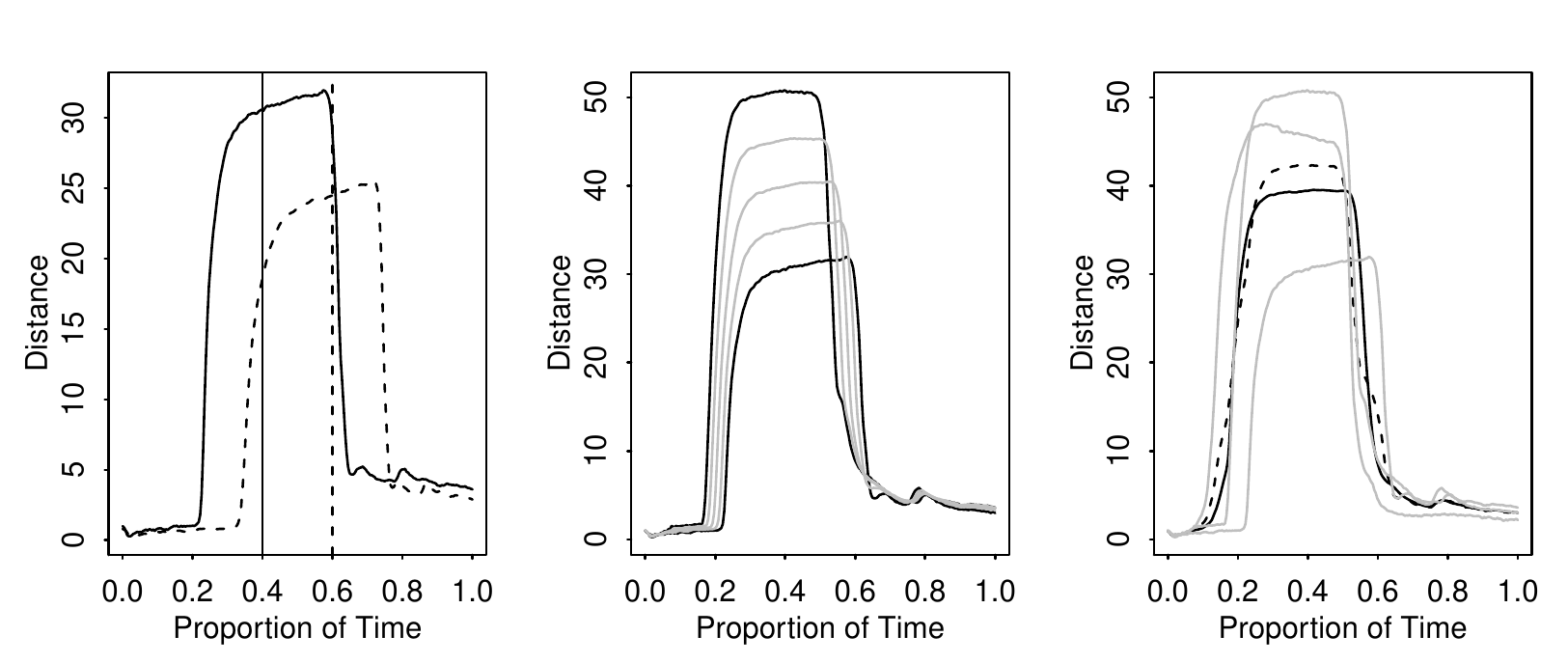}
  \caption{The first panel shows a transformation of an observed curve from solid to dashed. The second panel shows three curves (gray) along a geodesic linking two observed curves (black). The third panel shows the average (solid black) of three curves (gray) with the naive average shown as a dashed line.
  \label{fig:combfun}}
\end{figure}

We now demonstrate how to form the weighted combination of two functions $f_1$ and $f_2$. First choose $\mathbf{p^1}$ so that the distance $d(f_1^p,f_2)$ is minimized that is we transform $f_1$ to be close to $f_2$. We can define a sequence of functions between $f_1$ and $f_1^p$ by fixing $p_1$ but allowing the other two parameters to vary like this: $(p_1, p_1 + \gamma(p_2-p_1), 1 + \gamma(p_3-1))$ as $\gamma$ moves from 0 to 1. We can reverse the role of $f_1$ and $f_2$ to obtain $\mathbf{p^2}$.
We form the $\gamma$-weighted average by then combining $p_1$ and $p_2$ with the appropriate weights:
\begin{displaymath}
  \gamma f_1 + (1-\gamma) f_2 \equiv \gamma f_1^p(p_1^1,(1-\gamma) p_2^1+\gamma p_1^1, (1-\gamma)p_3^1+\gamma ) +
  (1-\gamma) f_2^p(p_1^2, \gamma p_2^2 + (1-\gamma) p_1^2 , \gamma p_3^2 + (1-\gamma) )
\end{displaymath}
As $\gamma$ varies from 1 to 0, it thus forms a kind of geodesic between $f_1$ and $f_2$. The second panel of Figure~\ref{fig:combface} shows weighted combinations shown as gray curves between the two solid observed curves for $\gamma = 0.25, 0.5, 0.75$.

We can form the weighted average of three functions by forming a weighted combination of the first two functions and then combining that with the third function with the appropriate weights. The lack of invariance to the ordering of the functions is inelegant but makes little difference in practice. An example is shown in the third panel of Figure~\ref{fig:combface}. We have shown the naive average $(f_1+f_2+f_3)/3$ for comparison. We can see that it has a lower slope on the leading edge than any of the three observed curves and an unexplained notch on the trailing edge of the curve. In contrast, the average computed using our method has similar properties to the observed curves. The more curves that we attempt to average in this manner, the greater the danger that the features will be attenuated and the qualitative smile property will be lost. We average over three curves at most to avoid this problem.

So now we have a distance on the curve space and a way to generate members of that space which is sufficient for us to proceed.
We compute the distance matrix for the curves and apply multidimensional scaling to this distance matrix. We extract the first two components as representative of the motion. As before, we use the tangent space coordinates for the shape space to represent the response as this is effectively equivalent to MDS on the distance matrix. Again we choose 10 components.  We then apply PLS as described earlier and find that two components are helpful in describing the relationship between the predictors and response.

We now describe how we obtain a mapping between the score space of $X$ and the original function space. To solve the optimization problem we first compute a centroid $\bar f$ using the following algorithm:
\begin{enumerate}
\item Order the functions by the distance of their score from $\mathbf{0}$. We pick the ten closest from which to build the mean. We could use all the functions but this increases the computation time to hours rather than minutes. Our choice means we restrict the search to functions which are likely to be close to the centroid.
\item For all triples of functions in the selected set of ten, find the weighted combination that has score $\mathbf{0}$ using Gower's scoring method. There may not be a solution for all triples.
\item For each feasible solution $f^*$, compute $\sum_i d(f^*,f_i)$. The estimated centroid function is the minimizing solution.
\end{enumerate}
This is not exactly the Fr\'echet median because we are not able to explore or define the space of feasible $f$ completely. Nevertheless, we do have a relatively dense exploration of the space so the approximation should be reasonable.

To obtain the best function corresponding to a target score of $s^*$:
\begin{enumerate}
\item Order the functions by the distance of their score from $s^*$. We pick the ten closest from which to build the solution.
\item For all triples of functions in this subset, find the weighted combination that has score $s^*$. Solutions will be found only for some triples.
\item For each feasible solution $f^*$, compute $d(f^*,\bar f)$. The estimated function is the minimizing solution.
\end{enumerate}
It is possible that for $s^*$ far from $\mathbf{0}$ that no solutions may found. It is in such cases that  allowing the weights $\gamma$ to vary  outside of $[0,1]$ (we use $[-0.1, 1.1]$) may be helpful although it simply may not be possible to estimate a function with too large a score. An alternative to taking the minimizing solution is to plot the estimates as this provides some notion of the variation among estimates with score $s^*$.

\begin{figure}[hbtp]
  \centering
  \includegraphics{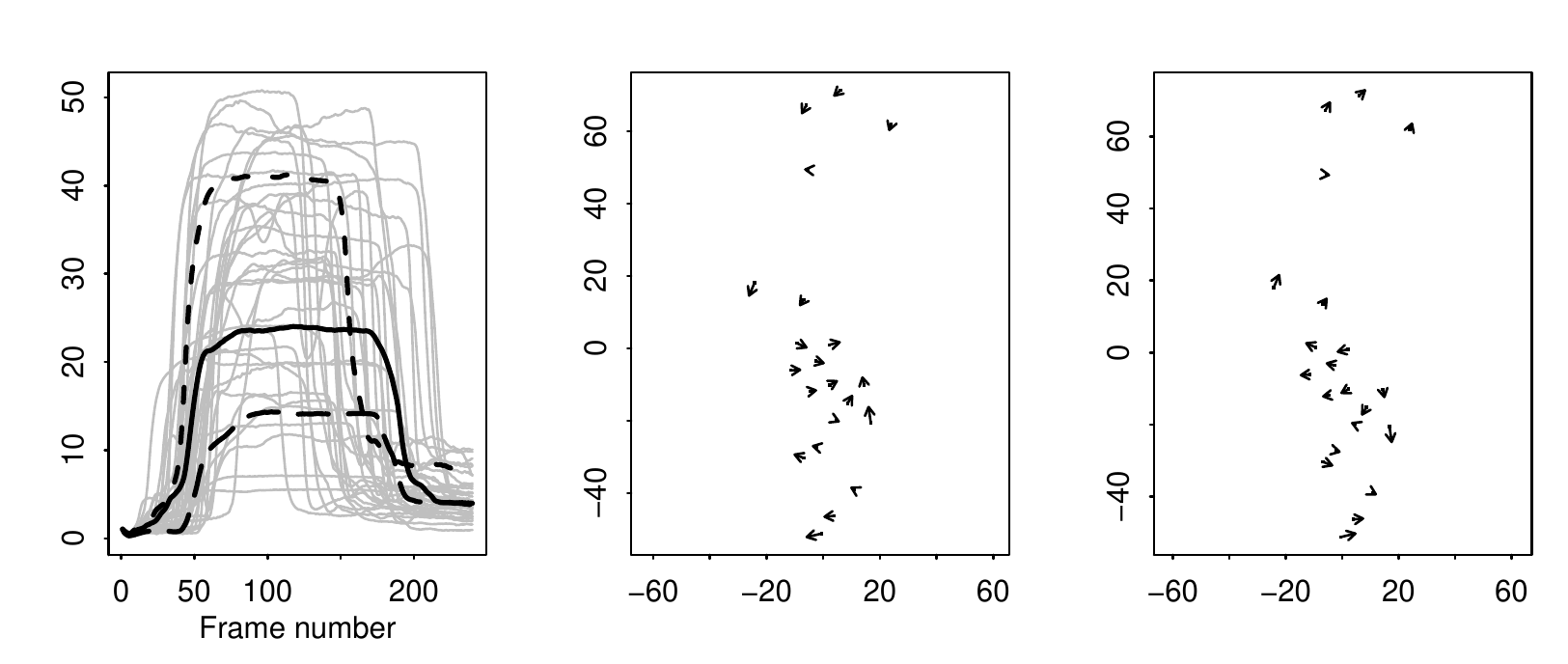}
  \caption{Same as Figure~\ref{fig:funcshap1} except for the second component of variation. The side rather than frontal view is shown. Here the second panel corresponds to the lower of the two dashed lines in the first panel while the third panel corresponds to the upper of the two.
  \label{fig:funcshap2}}
\end{figure}

We interpret the relationship using the following method. We compute the centroid in both the predictor and response space. For the predictor space, we consider each substantial component of variation, finding the predictor values corresponding to one SD perturbation around the centroid along the axis of this component of variation. For these two predictor values, we compute the corresponding predicted responses. For this example, two components are of interest shown in Figures~\ref{fig:funcshap1} and \ref{fig:funcshap2}.

We see the first component of variation in Figure~\ref{fig:funcshap1}. The lower of the two dashed lines in the first panel represents a one SD perturbation from the centroid (shown as the solid curve). This curve corresponds with a smaller smile of lesser duration. The second panel shows the response shape associated with this perturbation which describes a lips-together smile. The upper of the two dashed lines, representing the perturbation of one SD along the other direction of the first component, is a smile that is larger and lasts longer. The third panel is the response shape corresponding to this input which shows a mouth open smile.
So we see that a shorter, smaller smile is associated with a lips-together smile while the longer, larger smile is associated with a mouth-open smile.

The second component of variation is seen in Figure~\ref{fig:funcshap2}. In this case, more of the response variation is visible from the side of the face rather than the front of the face.  We see that a smaller, later smile (lower of the two dashed lines) is associated with a lips out, chin in smile (third panel) while a larger, sooner smile (upper of the two dashed lines) is associated with lips back, chin out smile (second panel).

We can compute a test of significance for the predictor in this regression. The test statistic has a value of 0.725. Using a permutation test where we permute the response, we determine that the $p$-value is very small (none of the 1000 permuted test-statistics exceeded the observed value). Although the effects are statistically significant, the size of the effect is smaller than that observed in the first example where initial pose was used as a predictor. In fact, the two response panels in Figures~\ref{fig:funcshap1} and \ref{fig:funcshap2} show perturbations multiplied by five as some amplification was necessary to make the effects clear. The actual effects are quite small.

We now explore the relative importance of other potential predictors. We use the initial pose, as described in the first example, the time-course function from earlier in this example and the centroid size of the initial face, combining these predictors as described in Section~\ref{stec:adding-predictors}. We compute the value of $R^2$ for all subsets of size two or more as seen in Table~\ref{tab:r2}. We see that the initial pose is the most important predictor of the maximal pose but we may question if the time and size variables are needed in addition to the initial pose.

\begin{table}[hbtp]
  \centering
\begin{tabular}{|ll|} \hline
  Model & $R^2$ \\ \hline
initial + time + size & 55.4 \\
time + size & 10.9 \\
initial + time & 54.6 \\
initial + size & 52.8 \\ \hline
\end{tabular}
  \caption{$R^2$ for models predicted maximal pose.
  \label{tab:r2}}
\end{table}

We test for the significance of the size variable using the F-statistic and assessing the significance using a permutation test that permutes only the size variable for each resample. We obtain a p-value of 0.58 from 1000 replications and conclude that the size does not have a statistically significant effect relative to this model. Further, we test the significance of both the time and size variables using the same technique. Here we find a p-value of 0.78, indicating that we may predict the maximal pose with the initial pose alone.

\section{Correlation Matrices}
\label{sec:correlation-matrices}

Correlation (or covariance) matrices can arise as variables in a regression analysis. Examples include radio communication as in \shortciteN{herdin05:_correl_mimo}, brain imaging as in \shortciteN{Dryden2009} and longitudinal data modeling as in \citeN{Daniels2002}. The space of positive semi-definite symmetric matrices is not Euclidean which poses difficulties for analysis. For example, the simple entrywise average of two correlation matrices may not itself be a correlation matrix. Although parametric modeling with Wishart distributions may be possible, this is difficult and we present a simpler, non-parametric approach here.

Given a sample of $m \times m$ correlation matrices $C_1, \dots , C_n$, we can compute a distance matrix $D_{ij}$ where
$i,j = 1, \dots ,n$. Various distances for correlation matrices have been proposed, for example, see \shortciteN{herdin05:_correl_mimo} but we shall use the Frobenius distance in the application to follow. Whether the correlation matrices appear as responses or predictors, we may apply the principal coordinates analysis as described above. However, useful interpretation requires that we be able to backscore to the correlation space once the internal modelling has occurred. We must first compute the Fr\'echet median:
\begin{displaymath}
  \bar C = \arg \min_{C} \sum_{i=1}^n d(C_i, C)
\end{displaymath}
The optimisation is not straightforward because the space is non-Euclidean so we take a different approach. We define the weighted combination of correlation matrices as
\begin{displaymath}
  C_\gamma = \arg \min_{C} d(\sum_{i=1}^n \gamma_i C_i , C)
\end{displaymath}
which may be computed for the Frobenius norm using the algorithm due to \citeN{higham2002computing}.  Note that
$\sum_{i=1}^n \gamma_i C_i$ may be a correlation matrix in which case the solution is $C$. If it is not, we find the closest
(in Frobenius norm) correlation matrix $C$.  The implementation of the backscoring method requires a method to search over weighted combinations of $C_i$. Given a new score $s$ we may use a Lagrange-like method to find:
\begin{displaymath}
  C(s) = \arg \min_{C_\gamma} \{ d(\bar C, C_\gamma) + \delta ||\phi(C_\lambda)-s|| \}
\end{displaymath}
where $\delta$ is chosen in a balanced manner to allow minimisation of the first argument while enforcing the constraint of the second. \shortciteN{Dryden2009} discuss the use of other metrics which avoid the projection used here to work with the Frobenius norm but would be more difficult to implement for this purpose.

We illustrate regression modelling using correlation matrices using the data derived from the same experiment. Muscle coordination during facial motion is important in expression. In the study of patients with cleft lip and palate repairs, researchers would like to discover factors that affect this coordination. We derive a 38x38 correlation matrix that describes this correlation. The next paragraph describes how this correlation was computed. The details of this computation are not central to this example and may be skipped.

Markers on the face move in three dimensions, but in most cases, the motion is close to linear. We preprocess the data, first using generalised procrustes analysis to remove whole head motion and ordinary procrustes analysis to rotate the head onto a symmetrised face-forward configuration of the markers. The purpose of this step is to ensure that all 36 motions are face-forward without any whole head motion. For each of the 38 markers on the face, we have a 240x3 matrix describing the trajectory for each subject. We perform a location standardisation so that all these trajectories start from the origin. We then concatenate these location shifted matrices into a (240x36)x3 matrix and perform a principal components analysis. The first principal component identifies the major axis of motion for that marker. We compute the scores corresponding to this component which show the timing and magnitude of motion along that axis.  Hence for each subject, we may produce a 240x38 matrix of scores from which we may compute a correlation matrix. The correlations between the markers represent the extent to which the motions of the markers are correlated. We considered computing correlations directly on the 240x(38x3) observed motion matrices but the cross-correlation between the three coordinate directions would have been difficult to interpret.

We considered two covariates as possible predictors of these correlation matrices --- size and speed. We use the centroid size (which is related to the age of the child). We computed the Procrustes distance of each frame of motion from the initial frame of motion and use the maximum rate of change in this distance (slightly smoothed) as an estimate of speed. We regressed the first two coordinates of the PCO on the size and speed and found that only the speed was a significant predictor of the first principal coordinate.

\begin{figure}[hbtp]
  \centering
  \includegraphics{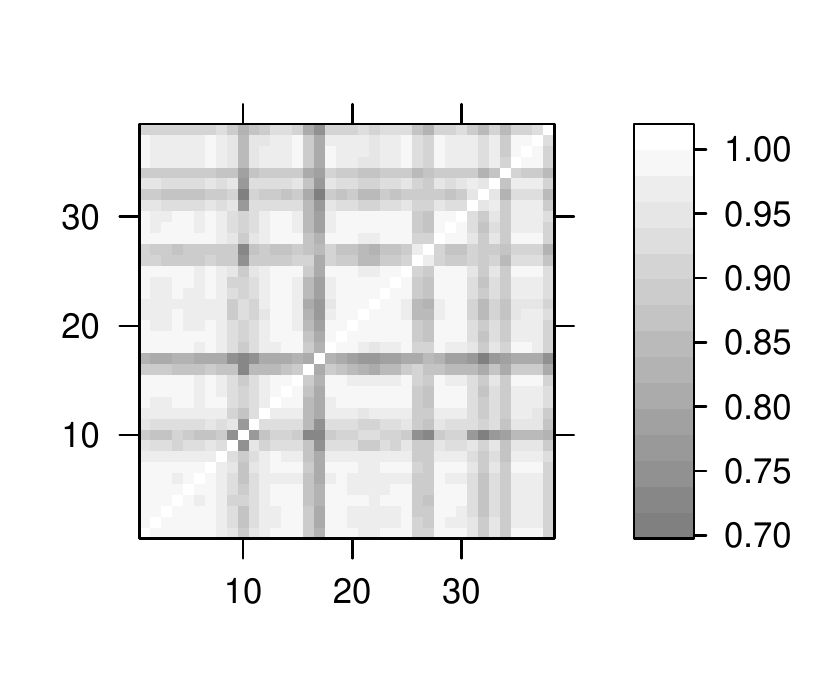}
  \includegraphics{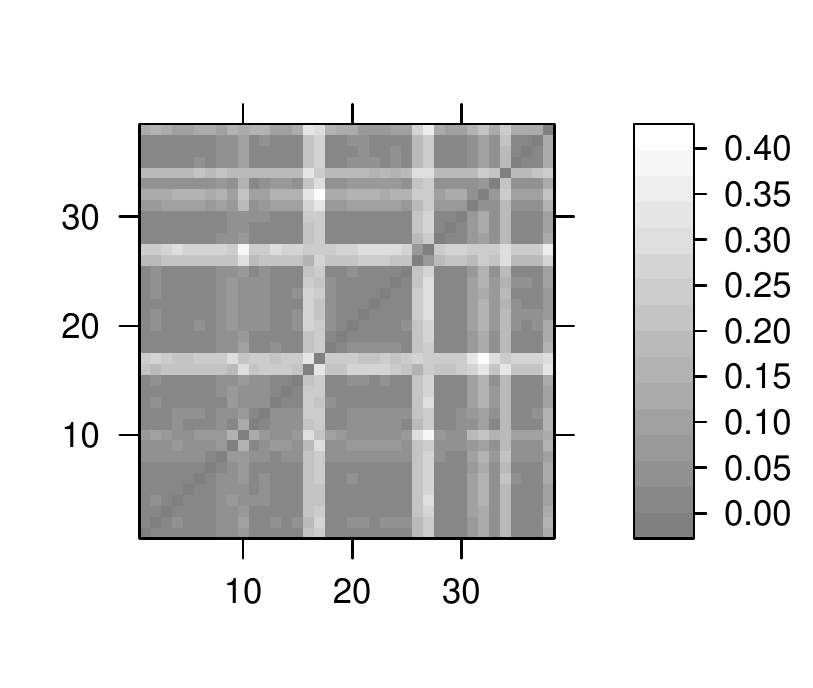}
  \caption{A heatmap of the mean correlation matrix is shown on the left. A heatmap of the difference between the predicted correlation for the maximum speed with the correlation for the minimum observed speed is shown on the right.
  \label{fig:corrplot}}
\end{figure}

Now consider an internal model regressing the first principal component on the speed alone. We contrast the predicted response for the minimum observed speed with the maximum observed speed. The predicted scores can be used to construct predicted correlation matrices. In Figure~\ref{fig:corrplot}, we can see the estimated mean correlation matrix which reveals many of the observed correlations are large but with some exceptions. The contrast in the correlations corresponding to the maximum and minimum observed speeds show that this predictor has little effect for many of the markers but that for markers 16, 17, 26 and 27 which are situated on the upper lip just below the nose (see Figure~\ref{fig:stdface}) and for markers 32,34 and 38 which are found on the jaw, there is a more substantial effect due to speed. At lower speeds, there is less coordination with other markers for these particular markers while at the higher speeds, these markers move more in conjunction with the other markers.

The traditional approach to interpreting scores from principal coordinates analysis (or classical multidimensional scaling) is to examine observed values with scores close to those of interest and then try to infer the meaning from these observations. In this case, we can compare the observed correlations closest to minimum and maximum speeds. However, these are single observations and subject to other unrelated kinds of variation. Such a contrast does not reveal the clear meaning shown in the second panel of Figure~\ref{fig:corrplot}.

\section{Discussion}
\label{sec:conclusion}

We have shown how data that lie in a non-Euclidean space may be integrated into a regression framework using only a concept of distance and limited assumptions about that space. The ability to backscore from the score space to the data space is essential for the method to be useful for prediction and explanation. The same idea may be expanded to other types of data such as images, tensors, trees and data of mixed types where at least a distance may be defined. This is in the spirit of \citeN{wang2007ood} who used the term ``object-oriented data analysis'' and developed some statistical methods for tree data types.

The mode of presentation in this paper has been a case study which has allowed us to demonstrate that the methodology can produce interesting and interpretable results. Nevertheless much additional work is necessary before the method can be used in practice with complete confidence. Theoretical questions regarding the properties of the backscoring method remain to be answered. The backscoring method is computationally intensive even for these small datasets so some improvement is necessary particularly if larger datasets are to be considered. Furthermore, the method of generating acceptable candidates during backscoring requires customization depending on the particular application. Some work is necessary to make these easier to generate, especially for common situations, would make the method more convenient to apply generally.

\bibliography{jfpapers}
\bibliographystyle{chicago}

\end{document}